\documentstyle[12pt]{article}

\begin{document}
\vskip 0.25in
\begin{center}
{\bf \LARGE  The effects of the massless
 $O(\alpha_s^2)$, $O(\alpha\alpha_s)$, $O(\alpha^2)$  QCD and QED
corrections   and of the massive contributions
to  $\Gamma(           H^0\rightarrow b\overline{b})$}
\end{center}
\vskip 0.2in
\begin{center}
{\bf Andrei L.~Kataev}\footnote{On way   from CERN, Switzerland to
the Institute for Nuclear Research, Moscow 117312, Russia (address
after 1 December 1992) ; e-mails:  bitnet:kataev@cernvm; internet:
kataev@inucres.msk.su }
\\ {\em{Laboratoire de Physique Th\'eorique}}
{\small E}N{\large S}{\Large L}{\large A}P{\small P}, \footnote{URA
14-36 du CNRS, associ\'ee  a l'E.N.S. du Lyon, at au L.A.P.P.
(IN2P3-CNRS) d'Annecy-le-Vieux}\\ {\em Chemin de Bellevue, BP 110,
F-74941 Annecy-le-Vieux Cedex, France}
\\ and
\\ {\bf Victor T. Kim}
\footnote{On leave of absence from St.Petersburg Nuclear Physics
Institute, Gatchina 188350, Russia (address after 1 December 1992
; bitnet:vkim@cernvm; internet:kim@lnpi.spb.su}\\
{\em CERN, CH-1211 Geneva 23, Switzerland} \end{center}
\begin{center} {\bf \Large Abstract}
\end{center} \vskip 0.15in
We consider in detail various theoretical uncertainties of the
perturbative predictions for the decay width of $H^0\rightarrow
b\overline{b}$ process in the region $50\ GeV< M_H\leq 2M_W$. We
calculate the order $O(\alpha_s^2)$-contributions to the expression
for $\Gamma_{Hb\overline{b}}$ through the pole quark mass and
demonstrate that they are important for the elimination  of
the numerical difference between the corresponding expression and the
one through the running $b$-quark mass. The order
$O(\alpha\alpha_s)$ and $O(\alpha^2)$ massless and order
$O(m_b^2/M_H^2)$ massive corrections to $\Gamma_{Hb\overline{b}}$
are also calculated. The importance of the latter contributions
for modeling of the threshold effects is
demonstrated. The troubles with identifying of
the 4 recent L3 events $e^+e^-\rightarrow
l^+l^-\gamma\gamma$ with the decay of a Standard
Higgs boson are discussed.


\addtocounter{page}{-1}
\thispagestyle{empty}
\vfill\eject
\pagestyle{empty}
\clearpage\mbox{}\clearpage
\pagestyle{plain}
\setcounter{page}{1}

\newpage
\setcounter{equation}{0}

{\bf 1.}~~~Among the most intriguing modern theoretical
 problems are the investigations of the properties of
still nondiscovered Higgs particles. The special attention is
nowadays paid to the searches of a Standard Model Higgs boson.
Various aspects of "Higgs hunting" were discussed in detail from
the theoretical \cite{Veltman},\cite{HHG} and the phenomenological
 \cite{HHG},\cite{Kunzt}, \cite{Dawson} points of view.

The current lower bound on a Standard Model Higgs boson mass is
$M_H>52\ GeV$ at the $95\%$ confidence level (see e.g. \cite{LEP}).
This lower bound came from the
 analysis of the LEP data with taking
into account of the theoretical expression for $H^0\rightarrow q\overline{q}$
decay rate including the results of the calculations of the order
$O(\alpha_s)$ QCD corrections \cite{Braaten}.
One of the main decay channels of a Higgs boson in the intermediate
mass range  $50\ GeV< M_H < 2M_W$ is the decay to the $b\overline{b}$
final states with the coupling constant being proportional to the
$b$-quark mass. In this mass range the theoretical uncertainties of
$\Gamma_{Hb\overline{b}}$=$\Gamma(H^0\rightarrow b \overline{b})$
 are closely related to the theoretical uncertainties of the
branching ratio of the process  $H^0\rightarrow\gamma
\gamma$ which is known as one of the most typical reactions
for the searches of not too heavy Higgs bosons.

Careful studies of various kinds of theoretical predictions for the
cross-sections and branching ratios of a Higgs particles became
 important in view of the appearance of the announced  by L3
collaboration 4 yet non-explained recent LEP events
in the reaction $e^+e^-\rightarrow l^+l^-\gamma\gamma$ (one
$e^+e^-\gamma\gamma$ and three $\mu^+\mu^-\gamma\gamma$ events) with
the invariant mass of the photons close to 60 GeV (!?) \cite{events}.
Moreover the analysis of the QCD uncertainties for
$\Gamma_{Hb\overline{b}}$ ,
  $Br(H^0\rightarrow\gamma\gamma)$ and for other branching ratios
can be useful in the planned searches of  a Higgs particles at the
 future colliders, namely LEP2, LHC, SSC.

The massless
 QCD corrections to $\Gamma(H^0\rightarrow hadrons)$ were considered
   at the next-to-next-to-leading order
(NNLO) using the concept of the running $b$-quark mass
$\overline{m}_b(M_H)$ \cite{GKL},\cite{Higgs} with taking into
account of the 3-loop NNLO corrections to the QCD
$\beta$-function \cite{TVZ} and to the anomalous mass dimension
 function \cite{Tarasov}.  In the process of the consideration of the
phenomenological consequences of the results of ref.\cite{Higgs} it
was noticed further on that taking into account of the variation of
 the running $b$-quark mass from the $b$-quark mass-shell
(namely from the $\overline{m}_b(m_b)$-value) to the
$M_H$-scale (namely to the $\overline{m}_b(M_H)$-value)
in the leading order (LO) of perturbation theory results in
the negative corrections, which diminish the value of the
corresponding Born expression for $\Gamma_{Hb\overline{b}}$ by over
$50\%$ \cite{Kunzt}.

However, the running mass is not the unique way of defining mass
parameters. Indeed, in total analogy with the physical
mass of electron in QCD one can
also define the pole quark mass
(for the general
consideration of the corresponding renormalization group equations
see ref.\cite{Shirkov}). The definition of the pole quark mass is
commonly used for heavy quarks, namely for $c$- and $b$-quarks.  The
expression for the decay width $H^0\rightarrow q\overline{q}$
($H^0\rightarrow l^-l^+$) has been explicitly calculated in terms of
the pole quark masses $m_c$ and $m_b$ in refs.\cite{Braaten},
\cite{Bardin} at the $O(\alpha_s)$ ($O(\alpha)$)-level. The presented
in ref.\cite{HHG} numerical studies of these results  did not reveal
the effect of the $50\%$ reduction of the Born approximation.
Therefore, it is important to understand the origin of the observed in
ref.\cite{HHG} puzzle of the differences between various
parametrizations of the QCD results for $\Gamma_{H b\overline{b}}$ in
the experimentally interesting region of $M_H$ values.

In this work using the
2-loop relation between the running and the pole quark masses
\cite{David} and the results of refs.\cite{Higgs}
we calculate the  expression for  $\Gamma_{Hb\overline{b}}$
 at the $\alpha_s^2$-level
in terms of the pole quark  mass in two different forms. The
first one will contain the $ln(M_H^2/m_b^2)$-contributions
explicitly, while in the second one they will be summed up through the
renormalization group (RG) technique. Note, that the second
parametrization is closely related to the one through the running
quark mass.

Our results demonstrate that for the RG-nonimproved expression
for $\Gamma_{H b\overline{b}}$ the calculated by us
$\alpha_s^2$-contribution  produce the
negative correction which is responsible for the elimination of
the numerical
difference between various parametrizations of
 the QCD results for
$\Gamma_{Hb\overline{b}}$.

We also present the more detailed analysis of the RG-improved (running
mass) expression for $\Gamma_{Hb\overline{b}}$
with taking into account of the effects, neglected in the course of
estimates of ref.\cite{Kunzt}. Among them are the corrections
responsible for the relation between  the running
mass $\overline{m}_b(m_b)$ and the pole mass $m_b$ at the
1-loop and 2-loop levels and the order $O(m_b^2/M_H^2)$-corrections to
 $\Gamma_{Hb\overline{b}}$.  The importance of taking into account of
the order $O(m_b^2/M_H^2)$-corrections for modeling the threshold
effects in the $H^0\rightarrow q\overline{q}$ process is
demonstrated. Following the lines of ref.\cite{AK} we also calculate
order $O(\alpha^2)$ and $O(\alpha\alpha_s)$ corrections to
$\Gamma_{Hb\overline{b}}$ and demonstrate that they are small.

Our pole mass dependent results  support the effect of
the $50\%$ reduction of the value of $\Gamma_{Hb\overline{b}}$
making it more theoretically substantiated. As to the phenomenology,
we discuss the troubles of the relation of
the 4 events announced by L3 group \cite{events}
with the decay of the standard Higgs boson with the mass
$M_H=60\ GeV$.

{\bf 2.}~~~We start from technical considerations. First we
remind different QCD expressions for
$\Gamma_{H b\overline{b}}$ in terms of the pole
quark mass which is defined as the position of the pole of the quark
 propagator. At the Born level of perturbation theory one has
\begin{equation} \Gamma_{H
b\overline{b}}=\Gamma(H^0\rightarrow{b\overline{b}})=
\Gamma_0^{(b)}\beta^3 \label{1} \end{equation}
where \begin{eqnarray}
\Gamma_{0}^{(b)} &=& \frac{3 \sqrt 2}{8\pi}G_F M_H m_b^2 \label{2} \\
\beta &=& \sqrt{1-\frac{4m_b^2}{M_H^2}} \nonumber
\end{eqnarray} The corresponding $O(\alpha_s)$ leading order (LO)
approximation of the decay width reads \cite{Braaten,Bardin}
\begin{equation} \Gamma_{Hb\overline{b}}=
\Gamma_0^{(b)} \beta^3\bigg[1+\delta\frac{\alpha_s(M_H)} {\pi}\bigg]
\label{3} \end{equation}
where \begin{equation}
\delta=
\frac{4}{3}\bigg[\frac{A(\beta)}{\beta}+\frac{3+34\beta^2-13\beta^4
}{16\beta^3}\ln{\frac{1+\beta}{1-\beta}}
+\frac{3(-1+7\beta^2)}{8\beta^2}\bigg]
\label{4}
\end{equation}
with $A(\beta)$ defined as
\begin{eqnarray}
A(\beta) &=& (1+\beta^2)\bigg[4Li_2\bigg(\frac{1-\beta}{1+\beta}\bigg)
+2Li_2\bigg(-\frac{1-\beta}{1+\beta}\bigg)
-3 \ln \frac{2}{1+\beta}\ln\frac{1+\beta}{1-\beta}-
2\ln\frac{1+\beta}{1-\beta}\ln\beta\bigg]
\nonumber \\
  &-& 3\beta \ln \frac{4}{1-\beta^2}-4\beta \ln \beta
\label{5}
\end{eqnarray}
Here $Li_2(x)=\int_0^x(dt/t)\ln(1-t)$ is the Spence function
\footnote{Notice that the expression for $A(\beta)$ in the second work
from ref.\cite{Bardin} contained misprint-last two terms of eq.(5)
were absent.}.

It is well-known that
in the limit $m_b^2/M_H^2\rightarrow 0$ (or $\beta\rightarrow 1$) the
expression for the $\alpha_s$-correction to $\Gamma_{Hb\overline{b}}$
appears to be infinite, namely it contains the logarithmic
singularities \begin{equation}
\Gamma_{Hb\overline{b}}=\Gamma_0^{b}\bigg[[1+\bigg(3-2ln(x)\bigg)
\frac{\alpha_s(M_H)}{\pi}]-6\frac{m_b^2}{M_H^2}[1+\bigg(\frac{4}{3}-
4ln(x)\bigg)\frac{\alpha_s(M_H)}{\pi}]+O(\frac{m_b^4}{M_H^4})\bigg]
\label{sing}
\end{equation}
where $x=M_H^2/m_b^2$. These logarithmic terms are connected to the
renormalization of the overall Yukawa coupling, namely to  the
$b$-quark mass, which appear in the expression for
$\Gamma_{Hb\overline{b}}$ at the leading order level. In order to
avoid these logarithmic terms one can sum them up by replacing the
pole  $b$-quark mass by the normalized on the $M_H$-scale running
$b$-quark mass $\overline{m}_b(M_H)$. Both the running quark mass and
the pole  quark mass are commonly used in QCD for the parametrization
of the massive-dependent contributions to physical quantities. The
relation between these two definitions of  quark masses will be
discussed in the next Section.

We will now consider the expression for $\Gamma_{Hb\overline{b}}$ in
terms of  $\overline{m}_b=\overline{m}_b(M_H)$
with taking into account of the order
$O(\overline{m}_b^2/M_H^2)$-contributions.  Since
$\overline{m}_b(M_H)$ is defined through the running coupling constant
$\alpha_s(M_H)$ already in the lowest order of perturbation theory, we
will identify the Born RG-improved expression with the LO
approximation, which can be defined as
\begin{equation}
\Gamma_{Hb\overline{b}}=\Gamma_0^{(b)}\frac{\overline{m}_b^2}{m_b^2}
\bigg(1-\frac{4\overline{m}_b^2}{M_H^2}\bigg)^{\frac{3}{2}}
\label{6}
\end{equation}
where in the LO of perturbation theory
                $\overline{m}_b(M_H)$=$(\alpha_s(M_H)/\alpha_s(m_b)
)^{12/23}  m_b$.
At the  NNLO level of
 perturbation theory the expanded in
$O(\overline{m}_b^2/M_H^2)$ -terms expression for
$\Gamma_{Hb\overline{b}}$ can be presented in the following form
\begin{equation}
\Gamma_{Hb\overline{b}}=\Gamma_0^{(b)}\frac{\overline{m}_b^2}{m_b^2}
\bigg[ \big[1+\Delta\overline{\Gamma}_1+
\Delta\overline{\Gamma}_2\big]
-6\frac{\overline{m}_b^2}{M_H^2}
\big[1+\Delta\overline{\Gamma}_1^{(m)}+\Delta\overline{\Gamma}_2^{(m)}
\big]+O\bigg(\frac{\overline{m}_b^4}{ M_H^4}\bigg)\bigg]
\label{7}
\end{equation}
Here $\Delta\overline{\Gamma}_1$ ($\Delta\overline{\Gamma}_1^{(m)}$)
and $\Delta\overline{\Gamma}_2$ ($\Delta\overline{\Gamma}_2^{(m)}$)
are the
next-to-leading order (NLO) and NNLO QCD corrections to
the "coefficient functions" of the $H^0\rightarrow b\overline{b}$
decay width $\Gamma_{Hb\overline{b}}$. The
general analytic expressions for $\Delta\overline{\Gamma}_1$ and
$\Delta\overline{\Gamma}_2$ were obtained in ref. \cite{Higgs}
in the $\overline{MS}$-scheme with the help of the SCHOONSHIP
analytical system \cite{SCH}.
In the case of QCD with $f=5$ numbers of flavours they have the
following numerical form \cite{Higgs} \begin{equation}
\Delta\overline{\Gamma}_1=\frac{17}{3} \frac{\alpha_s(M_H)}{\pi}
\label{10} \end{equation} \begin{equation}
\Delta\overline{\Gamma}_2=29.14
\bigg(\frac{\alpha_s(M_H)}{\pi}\bigg)^2 \label{11} \end{equation}
The result for $\Delta\overline{\Gamma}_1$ coincides with the one
obtained previously \cite{Sakai}. The  expression  for
$\Delta\overline{\Gamma}_2$  was recently confirmed
\cite{Oleg}
using FORM analytical system \cite{FORM}.

Let us determine the  NLO $\alpha_s$ correction
 to the order $O(\overline{m}_b^2/M_H^2)$
-contributions  using the results of calculations
of ref. \cite{Fyodor}. In the $\overline{MS}$-scheme it reads
\begin{equation}
\Delta\overline{\Gamma}_1^{(m)}=5 C_F\frac{\alpha_s(M_H)}{\pi}
=(C_F=4/3)=\frac{20}{3}\frac{\alpha_s(M_H)}{\pi}
\label{12}
\end{equation}
The corresponding NNLO
 $\alpha_s^2$-term is still unknown. In principle it
can be obtained after the analysis of the
 results of the calculations of the 3-loop order
$O(\overline{m}^2)$-
corrections to the 2-point function of the scalar quark currents
in the eucledian region \cite{s} with taking into account
of the effects of the analytical continuation to the physical region
($\pi^2$-terms). However,
 varying this correction between $\Delta\overline{\Gamma}_2^{(m)}$=0
and $\Delta\overline{\Gamma}_2^{(m)}$=40$(\alpha_s(M_H)/\pi)^2$
we will demonstrate that this term is not
very important in our phenomenological studies.

{\bf 3.}~~~Let us now express the corresponding approximation for
$\Gamma_{Hb\overline{b}}$ of eq.(8) with the radiative corrections
defined by eqs.(9,10,11) through the b-quark pole mass.
We will use the obtained in  ref.\cite{David}
2-loop relation between the normalized on the mass-shell running
 $b$-quark mass in the $\overline{MS}$-scheme and the pole
 $b$-quark mass, namely
\begin{equation}
\overline{m}_b^2(m_b)=m_b^2\bigg[1-\frac{8}{3}\frac{\alpha_s(m_b)}{\pi}
-\bigg(26.89-2.08\sum_{f=u}^{c}(1-\frac{m_f}{m_b})\bigg)\bigg(
\frac{\alpha_s(m_b)}{\pi}\bigg)^2\bigg]
\label{david}
\end{equation}
In the rough estimates presented in ref.\cite{Kunzt} the corrections
of order $\alpha_s$ and $\alpha_s^2$ in eq.(\ref{david}) were
neglected. However, in view of  the sizeable value of $\alpha_s(m_b)$
it is desirable to take these corrections into account in the
detailed studies.  Using eq.(\ref{david}) we get the following 2-loop
relation between the $\overline{MS}$-scheme running and pole b-quark
masses :  \begin{equation}
\overline{m}_b^2(M_H)=m_b^2\bigg[1-\bigg(\frac{8}{3}+2\ln(x)\bigg)
\frac{\alpha_s(M_H)}{\pi}-
\label{trans}
\end{equation} \[ -\bigg(26.89
-2.08\sum_{f=u}^{c}(1-\frac{m_f}{m_b})-\frac{1}{12}\ln^2(x)
+\frac{245}{36}\ln^2(x)\bigg)\bigg(\frac
{\alpha_s(M_H)}{\pi}\bigg)^2\bigg]
\]
The main phenomenological difference between the pole mass $m_b$ and
the running mass $\overline{m}_b(M_H)$ is that $m_b$ is the number
(over $m_b=4.6\ GeV$) usually extracted from the theoretical
studies of the properties of the $b$-quark bound states, while
$\overline{m}_b(s)$ is the asymptotically decreasing function,
which depends from the values of the QCD parameters
$m_b$ and $\Lambda_{\overline{MS}}$ (through the expression
for $\alpha_s$; see eq.(22) defined below).

 We will apply now eq.(13)  for the determination
of the higher order corrections to the expression of
$\Gamma_{Hb\overline{b}}$ through the pole mass $m_b$. The
corresponding  expression has the following form  \begin{equation}
\Gamma_{Hb\overline{b}}= \Gamma_0^{(b)}\bigg[\big[1+\Delta\Gamma_1
+\Delta\Gamma_2\big]-6\frac{m_b^2}{M_H^2}\big[
1+\Delta\Gamma_1^{(m)}+\Delta\Gamma_2^{(m)}\big]
+O\bigg(\frac{m_b^2}{M_H^4}\bigg) \bigg] \label{Gpole} \end{equation}
where \begin{eqnarray} \Delta\Gamma_1 &=&
\bigg(3-2 \ln(x)\bigg) \frac{\alpha_s(M_H)}{\pi} \\ \Delta\Gamma_2 &=&
\bigg(-4.54-2.08 \sum_{f=u}^{c}\frac{m_f}{m_b}
-18.14\ln(x)+0.08\ln^2(x)\bigg)\bigg(\frac{\alpha_s(M_H)}{\pi}\bigg)^2
 \\ \Delta\Gamma_1^{(m)} &=& \bigg(\frac{4}{3}-
4\ln(x)\bigg)\frac{\alpha_s(M_H)}{\pi} \\ \Delta\Gamma_2^{(m)} &=&
\Delta\overline{\Gamma}_2^{(m)}- \bigg(65.6
+4.16\sum_{f=u}^{c}\frac{m_f}{m_b} +29.6\ln(x)-4.17\ln^2
(x)\bigg)\bigg(\frac{\alpha_s(M_H)}{\pi}\bigg)^2 \end{eqnarray} The
results for $\Delta\Gamma_1$ and $\Delta\Gamma_1^{(m) }$ are in
 agreement with the ones obtained from the complete LO massive
 dependence of $\Gamma_{Hb\overline{b}}$ after expanding it in powers
of $m_b^2/M_H^2$ (see eq.(\ref{sing})).  The expressions for the order
$\alpha_s^2$ corrections are new.

In order to understand the relative value of the sizeable
$\ln(M_H^2/m_b^2)$-terms in the above presented parametrization of
$\Gamma_{Hb\overline{b}}$ we will sum them back to the running b-quark
mass $\overline{m}_b$ by solving the corresponding RG equation
\begin{equation}
\frac{\overline{m}_b^2(M_H)}{\overline{m}_b^2(m_b)}=
exp \bigg[-2\int_{\alpha_s(m_b)}^{\alpha_s(M_H)}
\frac{\gamma_m(x)}{\beta(x)}dx\bigg]
\label{RG}
\end{equation}
with taking into account of the 3-loop approximations of the QCD
$\beta$-function \cite{TVZ} and of the anomalous mass dimension
function $\gamma_m(\alpha_s)$ \cite{Tarasov}. Using then eq.(12) we
express $\overline{m}_b(M_H)$ through the pole quark mass $m_b$ as
\begin{equation}
\overline{m}_b^2(M_H)=m_b^2\Phi_b(\alpha_s(M_H),\alpha_s(m_b))
\label{relation}
\end{equation}
The NNLO approximation of the $\Phi_c$-function
in the $\overline{MS}$-scheme is determined by
the considerations of refs.\cite{Higgs,David}  and read
\begin{equation}
\Phi_b=\bigg(\frac{\alpha_s(M_H)}{\alpha_s(m_b)}\bigg)^
{\frac{24}{23}}\bigg[\frac{1+2.34\frac{\alpha_s(M_H)}{\pi}
+4.37\bigg(
\frac{\alpha_s(M_H)}{\pi}\bigg)^2}{1+2.34\frac{\alpha_s(m_b)}{\pi}
+4.37\bigg(\frac{\alpha_s(m_b)}{\pi}\bigg)^2}\bigg]\times
\label{phi}
\end{equation}
\[
\bigg
[1-2.67\frac{\alpha_s(m_b)}{\pi}-\bigg(18.58+2.08\sum_{f=u}^{c}\frac{m_f}
{m_b}\bigg)\bigg(\frac{\alpha_s(m_b)}{\pi}\bigg)^2\bigg]
\]
where we will use the following NNLO approximation of $\alpha_s$
in the $\overline{MS}$-scheme
\begin{equation}
\frac{\alpha_s(s)}{\pi}= \frac{1}{\beta_1\L_s}-\frac{\beta_2\ln\L_s}
{\beta_1^3\L_s^2}+\frac{1}{\beta_1^5\L_s^3}(\beta_2^2\ln^2\L_s-
\beta_2^2\ln\L_s+\beta_1\beta_3-\beta_2^2)
\label{alpha}
\end{equation}
for $f=5$ numbers of flavours
with $\L_s=\ln(s/\Lambda_{\overline{MS}}^2)$,  $\beta_1=23/12$,
$\beta_2=29/12$ and $\beta_3=9769/3456$ \cite{TVZ}.

Substituting now eq.(\ref{relation}) and eq.(\ref{phi}) into eq.(7)
and eq.(8) we get the LO RG-improved expression for
 $\Gamma_{Hb\overline{b}}$ through the b-quark pole mass
\begin{equation} \Gamma_{Hb\overline{b}}=\Gamma_0^{(b)}
                                \bigg(\frac{\alpha_s(M_H)}{\alpha_s(m_b)}
\bigg)^{\frac{24}{23}}\bigg(1-\frac{4m_b^2}{M_H^2}
\bigg(\frac{\alpha_s(M_H)}{\alpha_s(m_b)}\bigg)^{\frac{24}{23}}
\bigg)^{\frac{3}{2}}
\label{lo}
\end{equation}
and the corresponding NNLO expression
\begin{equation}
\Gamma_{Hb\overline{b}}=\Gamma_0^{(b)}
                                \bigg(\frac{\alpha_s(M_H)}{\alpha_s(m_b)}
\bigg)^{\frac{24}{23}}
\bigg[\big[1+\Delta\tilde{\Gamma}_1+\Delta\tilde{\Gamma}_2
\big]-6\frac{m_b^2}{M_H^2}\bigg(\frac{\alpha_s(M_H)}{\alpha_s(m_b)}\bigg)
^{\frac{24}{23}}
\big[1+\Delta\tilde{\Gamma}_1^{(m)}+\Delta\tilde{\Gamma}_2
^{(m)}\big]+O\bigg(\frac{m_b^4}{M_H^4}\bigg)\bigg]
\label{rg}
\end{equation}
where
\begin{eqnarray}
\Delta\tilde{\Gamma}_1 &=& 8.01\frac{\alpha_s(M_H)}{\pi}
-5.01\frac{\alpha_s(m_b)}{\pi}
\\
\Delta\tilde{\Gamma}_2 &=& 46.8\bigg(\frac{\alpha_s(M_H)}{\pi}\bigg)^2-
(11.2+2.08\sum_{f=u}^{c}\frac{m_f}{m_b})\bigg(\frac{\alpha_s(m_b)}{\pi}
\bigg)^2-40.13\frac{\alpha_s(M_H)}{\pi}\frac{\alpha_s(m_b)}{\pi}
\\
\Delta\tilde{\Gamma}_1^{(m)} &=& 11.35\frac{\alpha_s(M_H)}{\pi}
-10.02\frac{\alpha_s(m_b)}{\pi}
\end{eqnarray}
\begin{equation}
\Delta\tilde{\Gamma}_2^{(m)} = \Delta\overline{\Gamma}_2^{(m)}
+45.4\bigg(\frac{\alpha_s(M_H)}{\pi}\bigg)^2+(2.7-4.16\sum_{f=u}^{c}
\frac{m_f}{m_b})\bigg(\frac{\alpha_s(m_b)}{\pi}\bigg)^2
-113.7\frac{\alpha_s(M_H)}{\pi}\frac{\alpha_s(m_b)}{\pi}
\end{equation}
For the estimate of the numerical contribution of the
$\Delta\overline{\Gamma}_2^{(m)}$-term we will take
$\Delta\overline{\Gamma}_2^{(m)}$=0 and $\Delta\overline{\Gamma}_2^
{(m)}=40(\alpha_s(M_H)/\pi)^2$.

This considered by us parametrization of
$\Gamma_{Hb\overline{b}}$ is closely related to the one through the
running $b$-quark mass
$\overline{m}_b$ (see eq.(8)). The  difference results from
the truncation of the corresponding perturbative series at the NNLO
level after the substitution of eqs.(\ref{relation}),(\ref{phi}) into
eq.(8). We think that this parametrization is more convenient for the
study of the numerical difference between LO, NLO and NNLO RG-improved
parametrizations of $\Gamma_{Hb\overline{b}}$ (for the detailed
discussion see Sec.4) and for the analysis of the
theoretical uncertainties of the phenomenological estimates of
ref.\cite{Kunzt}.

{\bf 4.}~~~In order to understand the relative value of different
QCD corrections to $\Gamma_{Hb\overline{b}}$ discussed and derived
in Secs.2,3 we plot at Figs.1-4 the expression for the ratio
$R_{Hb\overline{b}}=\Gamma_{Hb\overline{b}}/\Gamma_0^{(b)}$
($\Gamma_0^{(b)}=3\sqrt{2}/(8\pi)G_FM_Hm_b^2$) as the functions of $M_H$
for the values of $\Lambda_{\overline{MS}}^{(5)}=150\ MeV$ and $m_b=
4.8\ GeV$ without and with order $m_b^2/M_H^2$-corrections.

Fig.1 corresponds to the massless expression of $R_{Hb\overline{b}}$
with $\Gamma_{Hb\overline{b}}$ defined through the pole quark mass
$m_b$ via eq.(14) with (a) $\Delta\Gamma_1=0$, $\Delta\Gamma_2=0$ ;
(b) $\Delta\Gamma_1$=(eq.(15)), $\Delta\Gamma_2 =0$ and (c)
$\Delta\Gamma_1$=(eq.(15)), $\Delta\Gamma_2$=(eq.(16)).

At Fig.2 the same ratio is expressed using the RG-improved massless
formula of for $\Gamma_{Hb\overline{b}}$ (see eq.(24)) with (a)
$\Delta\tilde{\Gamma}_1$=0 , $\Delta\tilde{\Gamma}_2$=0 ; (b)
$\Delta\tilde{\Gamma}_1$=(eq.(25)) , $\Delta\tilde{\Gamma}_2$=0 and
(c) $\Delta\tilde{\Gamma}_1$=(eq.(25)) ,
$\Delta\tilde{\Gamma}_2$=(eq.(26)) and the QCD coupling constant
$\alpha_s$ defined by the LO, NLO and NNLO approximations of eq.(22)
correspondingly.

Fig.3 shows the dependence of $R_{Hb\overline{b}}$ from $M_H$ in the
case of the pole quark mass parametrization of
$\Gamma_{Hb\overline{b}}$  with taking into account of the
$O(m_b^2/M_H^2)$-corrections. The solid curve (a) displays the
complete massive dependence of the Born approximation for
$R_{Hb\overline{b}}$ with $\Gamma_{Hb\overline{b}}$ defined by
eq.(1); the dashed curve (a) corresponds to the expanded in
$m_b^2/M_H^2$-term LO expression for $\Gamma_{Hb\overline{b}}$ (see
eq.(14) with $\Delta\Gamma_1$=0, $\Delta\Gamma_2$=0,
$\Delta\Gamma_1^{(m)}$=0 and $\Delta\Gamma_2^{(m)}$=0). The
dashed-dotted curve (b) demonstrates the dependence from $M_H$ of the
complete LO expression for $R_{Hb\overline{b}}$ with
$\Gamma_{Hb\overline{b}}$ defined by eqs.(3)-(5)); the dotted
curve (b) corresponds to the LO approximation of
$\Gamma_{Hb\overline{b}}$ (see eq.(14)) with
$\Delta\Gamma_1$=(eq.(15)), $\Delta\Gamma_2$=0,
$\Delta\Gamma_1^{(m)}$=(eq.(17)), $\Delta\Gamma_2^{(m)}$=0. The
 curves (c) compare the NLO behavior of the expanded expression for
 $R_{Hb\overline{b}}$ with $\Gamma_{Hb\overline{b}}$ defined by
 eqs.(14)-(18) with $\Delta\overline{\Gamma}_2^{(m)}$=0 (solid
 curve) and
 $\Delta\overline{\Gamma}_2^{(m)}$=40$(\alpha_s(M_H)/\pi)^2$ (dashed
 curve) .

The similar dependence of the RG-improved results of eqs.(23),(24)
are depicted at Fig.4. The solid curve (a) corresponds to the RG-improved
 non-expanded LO massive dependent expression for $R_{Hb\overline{b}}$
where $\Gamma_{Hb\overline{b}}$ is defined by
eq.(23) with the LO approximation of $\alpha_s$ ; the dashed curve
 (a) demonstrates the behavior of the expanded in masses LO
RG-improved expression for $R_{Hb\overline{b}}$ with
$\Gamma_{Hb\overline{b}}$ defined by eq.(24) with
$\Delta\tilde{\Gamma}_1$=0, $\Delta\tilde{\Gamma}_2$=0,
$\Delta\tilde{\Gamma}_1^{(m)}$=0 , $\Delta\tilde{\Gamma}_2^{(m)}$=0
 and the LO approximation of $\alpha_s$.  The dotted curve (b) gives
the understanding of the behavior of the NLO RG-improved approximant
for $R_{Hb\overline{b}}$ with $\Gamma_{Hb\overline{b}}$ defined by
eq.(24) with $\Delta\tilde{\Gamma}_1$=(eq.(25)),
$\Delta\tilde{\Gamma}_2$=0, $\Delta\tilde{\Gamma}_1^{(m)}$=(eq.(27)),
$\Delta\tilde{\Gamma}_2^{(m)}$=0 and the NLO approximation of
$\alpha_s$. Finally  the curves (c) demonstrate the behavior of the
NNLO RG-improved massive approximation of $R_{Hb\overline{b}}$ defined
by eqs.(24)-(28) with $\Delta\overline{\Gamma}_2^{(m)}$=0 (solid
 curve) and
 $\Delta\overline{\Gamma}_2^{(m)}$=40$(\alpha_s(M_H)/\pi)^2$ (dashed
 curve).

Let us now discuss our understanding of the behavior of the different
approximations of $R_{Hb\overline{b}}$ for
$\Lambda_{\overline{MS}}^{(5)}=150\ MeV$.
\begin{enumerate}
\item For the pole mass parametrization of eq.(3) and eq.(14)
taking into account of the LO and NLO QCD corrections far
above threshold region decreases the value of $R_{Hb\overline{b}}$ by
over $40\%$ and $10\%$ correspondingly in both massless and
massive-dependent cases (see Figs.1,3). At the NLO level the
total numerical reduction is therefore of over $50\%$.
\item For the RG-improved parametrization the similar pattern can
be observed already
 for the LO Born results of eq.(23),(24) (see Figs.2,4).
This welcomed feature, which was already noticed in ref.\cite{Kunzt}
starting from the results of ref.\cite{Higgs},
is related to the effect of variation of the
running b-quark mass from the on-shell scale (namely from $\overline{
m}_b(m_b)$) to the $M_H$ scale (i.e. to $\overline{m}_b(M_H)$) and
demonstrate the importance of application of the RG-formalism in the
phenomenological studies.
\item Figs.2,4 show that, on the contrary to the pole mass approach,
in the RG-approach the LO approximation does not differ significantly
from the NLO and the NNLO results. This is the one more
welcomed feature of the RG-analysis, which demonstrates that the
theoretical uncertainties of the RG-improved Born approximation are
significantly smaller of the uncertainties of the Born expression for
$\Gamma_{Hb\overline{b}}$ through the pole quark mass.
Notice, however, that far above
the threshold region the NLO corrections are smaller (sometimes they are
even zero!) then the negative NNLO ones, which decrease the LO and NLO
approximations by $2\%$ to $4\%$
 \footnote {On the contrary to the
considerations of \cite{Chetyrkin1,Chetyrkin} we expect the
manifestation of the similar behavior of the RG-improved
massive-dependent contributions to $Z^0\rightarrow b\overline{b}$
decay.}. This pattern might indicate the possible problems of
applicability of the NNLO perturbative approximation of
$\Gamma_{Hb\overline{b}}$ (see eq.(24)) and of the related 2-loop
 mass relation of eq.(14) already discussed from this point of view in
the original work on the subject \cite{David}.  \item One of our main
results is that the calculated by us $\alpha_s^2$-corrections to the
pole-mass parametrization of both "massless" and the order
$m_b^2/M_H^2$ contributions
to $\Gamma_{Hb\overline{b}}$ (see eqs.(14,16,18)) decrease the
 difference between the pole-mass and RG-improved expressions for
$R_{Hb\overline{b}}$ (compare Figs.1,3 with Figs.2,4). Indeed, the
$15\%$ difference of the order $O(\alpha_s)$ approximations shrinks
to over $4\%$ at the $O(\alpha_s^2)$-level. Therefore, the observed
in ref.\cite{HHG} puzzle of the differences between RG-non-improved
and RG-improved parametrizations of $\Gamma_{Hb\overline{b}}$
can be resolved after taking into account of these effects.
We have checked that the similar situation also holds in the case of
the vector and axial contributions to $Z^0\rightarrow b\overline{b}$
decay, which were calculated at the NNLO in ref.\cite{Chetyrkin1}
and ref.\cite{Chetyrkin} correspondingly. We hope to return to the
more detailed consideration of this topic in future.
\item The comparison of Figs.1,3 and Figs.2,4 demonstrate that
starting from $M_H\geq60\ GeV$ the order $O(m_b^2/M_H^2)$-contributions
can be safely neglected both in the pole-mass and RG-improved approaches.
 However, beyond this region these corrections are very important for
modeling the threshold behavior of the corresponding approximants.
\item Indeed, the comparison of two curves (a) and (b) of Fig.3
and of two curves (a) of Fig.4 shows that near the threshold region
the Born and the LO
 massive dependence of the pole-mass expression
for $R_{Hb\overline{b}}$ (see eqs.(3,4,5)) and
the  Born LO RG-improved
massive dependence (see eq.(23)) are nicely approximated by the
corresponding expanded formulae with taking into account of the
leading order $m_b^2/M_H^2$ and $\overline{m}_b^2/M_H^2$-terms.
In view of this excellent agreement we can hope that this might
be the feature, which do not depend neither from the
order of perturbation theory, nor from the concrete process.
This observation joins the results of the more rigorous considerations
of the non-expanded and expanded expressions of the 2-loop self-energy
diagrams with different masses \cite{Andrey} in the understanding
of the necessity of taking into account of the massive corrections
in the problems connected with the phenomenological studies of the
threshold effects in different processes, including
$e^+e^-\rightarrow{hadrons}$ and $Z^0\rightarrow{hadrons}$,
which are now under consideration \cite{Pako}. This observation is
another new result of our work.
 \item The comparison of the threshold behavior of the pole-mass and
RG-improved massive-dependent expressions for $R_{Hb\overline{b}}$
with different values of the unknown at present order
$O(\alpha_s^2m_b^2/M_H^2)$ term $\Delta\overline{\Gamma}_2^{(m)}$
(compare solid curves (c) at Figs.3,4 with
$\Delta\overline{\Gamma}_2^{(m)}=0$ and the corresponding dashed
curves (c) with $\Delta\overline{\Gamma}_2^{(m)}$=40$(\alpha_s(M_H)/
\pi)^2$) demonstrate that this term is not very important in modeling
the threshold behavior of the corresponding 2-loop approximants.
However, in view of the discussions of the previous Subsection 4.6 it
might be of theoretical interest to calculate this term explicitly .
\end{enumerate}

Taking into account the discussed above
 QCD uncertainties and the current uncertainties of the value of the
parameter $\Lambda_{\overline{MS}}^{(5)}$ (for the detailed
discussions see ref.\cite{Altarelli}) we obtain the following
theoretical estimate for $\Gamma_{Hb\overline{b}}$ in the intermediate
region of $M_H$ values $50\ GeV <M_H< 160  \ GeV$ \begin{equation}
\Gamma_{Hb\overline{b}}=\bigg(0.55\div
0.45\bigg)\frac{3\sqrt{2}}{8\pi} G_FM_Hm_b^2 \label{h}
\end{equation}
This result contain less theoretical uncertainties then the
estimates of ref.\cite{Kunzt}.
Even more refined estimates can be obtained from Figs.1-4
using the concrete values of $M_H$-mass.

It should be mentioned, that for
$\Lambda_{\overline{MS}}^{(5)}$=$250\ MeV$ the relative value of the
NLO corrections to the RG-improoved expression of
$\Gamma_{Hb\overline{b}}$ in the typical region of $M_H$ values is
slightly different then in the case of
$\Lambda_{\overline{MS}}^{(5)}$=$150\ MeV$, being identically zero for
$M_H\approx 65\ GeV$ \cite{KKL}. \footnote{We are far away from
extracting from this observation any theoretical consequences}.

Note, that in our considerations of the order $O(\alpha_s^2)$ QCD
massive-dependent corrections to $\Gamma_{Hb\overline{b}}$ we did
not take into account the possible contributions of the diagrams
of Fig.5 with the b-quark and t-quark virtual loops. These diagrams
arise from the 3-loop ones of Fig.6 after the unitarity cut (a)
(another cut (b) gives the contribution to $H^0\rightarrow gg$
subprocess). In the limit $M_H>>2m_q$ (namely in the case of the
neglecting of the masses of the quarks propagating in the internal
loops) these diagrams are zero identically. However, they can give
some massive-dependent corrections, not studied in ref.  \cite{s}.
Indeed, as was shown in ref.\cite{KK}, the calculation of the
contribution of the similar diagrams to the axial channel of
$Z^0\rightarrow b\overline{b}$ decay results in the appearance of the
 sizeable $\alpha_s^2\ln(m_t/M_Z)$-terms. In our case we also expect
the appearance of the  massive correction of the similar origin
(namely with the top-mass dependence).  We think  that
this, not yet calculated contribution, will not change significantly
our result of eq.(\ref{h}).

In the region $50\ GeV<M_H<160\ GeV$ the result (29) for
$\Gamma_{Hb\overline{b}}$ is also not affected by the 1-loop
electroweak (EW) corrections, calculated in ref.\cite{Bardin}.
Indeed, for the analyzed values of a Higgs mass these negative EW
corrections are negligible (less then $-1\%$) \cite{Bardin}.  They
start to play the significant role in the region $M_{H} > 250\ GeV$
\cite{Bardin}, not considered in our work.  Another source of
theoretical uncertainties in the $\Gamma_{Hb\overline{b}}$-expression
comes from the higher order EW-corrections (which due to the smallness
of the 1-loop EW effects in the analyzed region of $M_H$-values should
be small) and from the mixed EW-QCD contributions. Following
the lines of ref.\cite{AK} and using the results of \cite{Higgs} we
 can calculate the part of these corrections, namely the contributions
to $\Gamma_{Hb\overline{b}}$ of the QED corrections of order
$O(\alpha^2)$ and $O(\alpha\alpha_s)$.

The QED corrected expression for $\Gamma_{Hb\overline{b}}$ reads
\begin{equation}
\Gamma_{Hb\overline{b}}=\Gamma_{Hb\overline{b}}^{QCD}+
\Gamma_{Hb\overline{b}}^{QED}
\end{equation}
where the QCD contributions were discussed in detail in Secs.2,3,4.
The QED contribution $\Gamma_{Hb\overline{b}}$ can be presented as
\begin{equation}
\Gamma_{Hb\overline{b}}=\Gamma_0^{(b)}\frac{\overline{m}_b^2}{m_b^2}
\bigg[\Delta\overline{\Gamma}_{1,QED}+\Delta\overline{\Gamma}_{2,QED}
-6\frac{\overline{m}_b^2}{M_H^2}\Delta\overline{\Gamma}_{1,QED}^{(m)}
+O\bigg(\frac{\overline{m}_b^4}{M_H^4}\bigg)\bigg]
\end{equation}
where $\overline{m}_b=\overline{m}_b(M_H)$ is  defined with taking
into account of the QED corrections to the anomalous mass dimension
function $\gamma_m$.
The  expressions for
the QED corrections to the ``coefficient functions" have the following
form
\begin{eqnarray} \Delta\overline{\Gamma}_{1,QED} &=&
\frac{17}{4}Q_b^2\frac{\alpha(M_H)} {\pi}=0.47\frac{\alpha(M_H)}{\pi}
\\ \Delta\overline{\Gamma}_{2,QED} &=& \bigg[\bigg(\frac{691}{64}
-\frac{9}{4}\zeta(3)-\frac{3\pi^2}{8}\bigg)Q_b^4   \nonumber \\ & &
-\bigg(\frac{65}{16}-\frac{163}{16}\zeta(3)-\frac{\pi^2}{12}
\bigg)Q_b^2\bigg(3\sum_{j=u}^{b}Q_j^2+N\bigg)\bigg]
\bigg(\frac{\alpha(M_H^2)}{\pi}\bigg)^2  \nonumber \\
& & +\bigg(\frac{691}{24}-6\zeta(3)-\pi^2\bigg)Q_b^2
\frac{\alpha(M_H)}{\pi}\frac{\alpha_s(M_H)}{\pi}
\\
&=&
6.7\bigg(\frac{\alpha(M_H)}{\pi}\bigg)^2+1.3\frac{\alpha(M_H)}{\pi}
\frac{\alpha_s(M_H)}{\pi} \nonumber \\
\Delta\overline{\Gamma}_{1,QED}^{(m)} &=&
5Q_b^2\frac{\alpha(M_H)}{\pi} =0.56\frac{\alpha(M_H)}{\pi}
\end{eqnarray} where $Q_j$ are the charges of u,d,s,c and b-quarks, N
is the number of leptons (N=3) and $\alpha(M_H)$ is the QED running
coupling constant in the $\overline{MS}$-scheme.
One can see that the higher order QED contributions, as calculated by
us, are very small.

{\bf 5.}~~~ The analysed by us corrections do not affect the
conclusion that there are the problems with the reltion of the
announced by L3 collaboration 4 events \cite{events} with
 the decay of a Standard Higgs boson
with the mass $M_H=60\ GeV$. Indeed, the number of
$H\rightarrow\gamma\gamma$ events at the detector with the acceptance
$A$ is \begin{equation} N=A\times L\times\sigma_{tot}\times
Br(H\rightarrow\gamma\gamma) \label{number} \end{equation} where $L$
is the integrated luminocity and $\sigma_{tot}$ is the total
cross-section of a Higgs production. For the process
$e^+e^-\rightarrow q\overline{q},H$ with $M_H=60\ GeV$ $\sigma_{tot}$
with taking into account of the initial state radiation is about $0.4\
pkb$ at $Z$-peak \cite{higi}. \footnote{The process $e^+e^-\rightarrow
q\overline{q},H$ was first considered in ref.\cite{Ioffe} and then
discussed in ref.\cite{Bjorken}.} This means
that for the process $e^+e^-\rightarrow l^-l^+,H$ $\sigma_{tot}$ is
about $0.04\ pkb$. For the L3 data \cite{events} the integrated
luminocity is 27 $1/pkb$. Since for the $l^-l^+,2\gamma$ channel the
$L3$ detector has the acceptance $A=0.85$ , one has \begin{equation}
N<10^{-3}   (!) \end{equation}  Moreover, for $M_H=60\ GeV$ one
should have the same number of the events for the process
$e^+e^-\rightarrow H\gamma\rightarrow 3\gamma$. Indeed, for $M_H=60\
GeV$ $\sigma_{tot}(e^+e^-\rightarrow H\gamma)$=
$\sigma_{tot}(e^+e^-\rightarrow l^+l^-H)$ (see e.g.\cite{HHG}).
Notice, that $e^+e^-\rightarrow 3\gamma$ L3 data \cite{L31} do not
contradict QED.

\vspace{1cm} \noindent
 \large {\bf Conclusions.} \normalsize
\vspace{0.5 cm}

In this work  we consider    in detail various QCD uncertainties
of the perturbative predictions for the decay width of the $H^0
\rightarrow b\overline{b}$ process, which is the dominant decay mode of a
Higgs boson in the intermediate mass range $50\ GeV<M_H<2M_W$. We
calculate the $\alpha_s^2$-contributions to the expression for
$\Gamma_{Hb\overline{b}}$ through the b-quark pole mass.
We
present two different parametrizations of the results for the decay
width  $\Gamma_{Hb\overline{b}}$ through the b-quark pole mass and
demonstrate  that the
calculated by us $\alpha_s^2\ln^2(M_H/m_b)$ and
$\alpha_s^2\ln(M_H/m_b)$ terms (which are responsible for the
variation of the running b-quark mass at the NNLO level)
are important since they are decreasing  the
difference between the corresponding approximation   for
$\Gamma_{Hb\overline{b}}$ and the one obtained after the summation of
these $\ln(M_H/m_b)$-terms using the RG-technique. We also
analyse in detail various
massive corrections of order $O(m_b^2/M_H^2)$ and demonstrate their
importance for the analysis of the threshold contributions to
$\Gamma_{Hb \overline{b}}$.

Far above threshold region the analyzed by us QCD corrections are
decreasing the value of $\Gamma_{Hb\overline{b}}$ by the factor of
over 0.5 (for $\Lambda_{\overline{MS}}^{(5)}=150\ MeV$) or over 0.45
(for $\Lambda_{\overline{MS}}^{(5)}=250\ MeV$). The reduction of the
value of $\Gamma_{Hb\overline{b}}$ is manifesting itself most
obviously for the RG-improved expressions which is connected
with the running mass parametrization. This fact demonstrates
the importance of using the RG-method in the phenomenological studies.

The discussed QCD contributions to $\Gamma_{Hb\overline{b}}$
are increasing
                        the values of the branching ratios of the
     decays
                                      $H^0\rightarrow l^-l^+$ and
$H^0\rightarrow\gamma\gamma$ by the factor of over 2.
 The consideration of the latter process
with taking into account of all existing theoretical information
about its decay width \cite{hpp},\cite{Zerwas} can be  of interest
from the point of view of planning of the         experiments aimed to
the searches of the  signals from
              a Higgs boson in the intermediate mass range. In future
we are
going to return to the detailed considerations of the QCD uncertaintes
for
   the       branching ratios of different decays of a Standard Model
Higgs boson.

In the process of our work we became aware of the work \cite{Kleiss},
             where the effects of the NNLO corrections to
$\Gamma_{Hb\overline{b}}$ \cite{Higgs} were also considered
using the running mass parametrization.  The certain
physical results of our more detailed studies are in qualitative
agreement with the presented in refs.\cite{Kleiss}
considerations, which however do not touch the analysed by
us problems. The proposed by us  RG-improoved
parametrization of the massive dependent effects (see ref.\cite{KKL}
and the discussions above) was also subsequently used in
ref.\cite{Kostya} for the analysis of the massive-dependent
contributions to the axial $Z$-boson decay rate. This analysis confirm
our qualitative expectations. However, it might be also of
theoretical interest the further  more detailed study of the
prescription dependence (or scheme dependence) of the perturbation
serious predicions for the massive-dependent contributions to physical
quantities, say to $Z\rightarrow b\overline{b}$.

\vspace{1cm} \noindent
\large {\bf  Acknowledgements} \normalsize
\vspace{0.5cm}

We are grateful to G. Kane for the fruitful conversations. One of
authors (A.K.) is also grateful to P. Aurenche and P. Sorba for the
kind hospitality in LAPP, where the preliminary version of this work
\cite{KKL} was completed. V.K. wishes to thank A. A. Vorobyov for
the support of this work.

\newpage
\parindent 0.0cm
{\bf Figure captions}

\vspace {1.5cm}
Fig.1: The pole-mass parametrization of the different approximations
for $R_{Hb\overline{b}}=\Gamma_{Hb\overline{b}}/\Gamma_o^{(b)}$
without massive-dependent corrections ($\Lambda_{\overline{MS}}^{(5)}
=150\ MeV$).

\vspace {1.0cm}
Fig.2: The RG-improved approximation of the different expressions
for $R_{Hb\overline{b}}$ without massive-dependent corrections
($\Lambda_{\overline{MS}}^{(5)}=150\ MeV$).

\vspace {1.0cm}
Fig.3: The pole-mass parametrization of the different approximations
of $R_{Hb\overline{b}}$ with the massive-dependent corrections (
$\Lambda_{\overline{MS}}^{(5)}=150\ MeV$).

\vspace {1.0cm}
Fig.4: The RG-improved approximations of the different expressions
for $R_{Hb\overline{b}}$ with the massive-dependent contributions
($\Lambda_{\overline{MS}}^{(5)}=150\ MeV$).


\vspace {1.0cm}
Fig.5: The diagram with the top-quark loop, which was not taken
into account in our considerations.

\vspace {1.0cm}
Fig.6: The 3-loop graph which gives different non-analysed
by us contributions to \hfill\break
 $\Gamma(H^0\rightarrow{hadrons})$ after the
unitarity cuts.

\end{document}